\documentclass{emulateapj} 

\shorttitle{Luminosity function in Coma} \shortauthors{Milne et al.}

\begin{document}

\title{The Faint End of the Luminosity Function in the Core of the Coma Cluster}

\author{Margaret L. Milne\altaffilmark{1}, Christopher J. Pritchet, Gregory B. 
Poole and Stephen D. J. Gwyn} \affil{Department of Physics \& Astronomy, 
University of Victoria, Victoria, BC V8P 1A1, Canada
\\ mlmilne@uvastro.phys.uvic.ca, pritchet@uvic.ca, 
gbpoole@uvastro.phys.uvic.ca, gwyn@uvastro.phys.uvic.ca}

\author{J J Kavelaars} \affil{Herzberg Institute of Astrophysics, 
National 
Research Council of Canada, 5071 West Saanich Road, Victoria, BC V9E 2E7, 
Canada \\ jj.kavelaars@nrc.gc.ca}

\author{William E. Harris} \affil{Department of Physics and Astronomy, McMaster 
University, Hamilton, ON L8S 4M1, Canada \\ harris@physics.mcmaster.ca}

\and

\author{David A. Hanes} \affil{Department of Physics, Queen's University, 
Kingston, ON K7L 3N6, Canada \\ hanes@astro.queensu.ca}

\altaffiltext{1}{Current address: Centre of the Universe, National Research 
Council of Canada, 5071 West Saanich Road, Victoria, BC V9E 2E7, Canada; 
Margaret.L.Milne@nrc-cnrc.gc.ca}

\begin{abstract} We present optical measurements of the faint end of the 
luminosity function in the core of the Coma cluster.  Dwarf galaxies are 
detected down to a limiting magnitude of $R \sim 25.75$ in images taken with the 
Hubble Space Telescope.  This represents the faintest determination of the Coma 
luminosity function to date.  With the assumption that errors due to cosmic 
variance are small, evidence is found for a steep faint end slope with 
$\alpha \lesssim -2$.  Such a value is expected in theories in which 
reionization and other feedback processes are dependent on density. 
\end{abstract}


\section{INTRODUCTION}

The galaxy luminosity function (LF) is a fundamental observational probe of 
galaxy formation and evolution. Defined as the number of galaxies per unit 
magnitude per unit area on the sky$^2$, the LF depends on both the initial 
density fluctuation spectrum and on baryonic processes such as cooling, star 
formation and supernova feedback.  Any theory attempting to explain how galaxies 
form and evolve must test its predictions against the observed shape of the LF.

\footnotetext[2]{More precisely, this is the definition of the luminosity {\it 
distribution}.  The luminosity {\it function} is the number of galaxies per unit 
magnitude per unit {\it volume}.  The term ``luminosity function'' is often used 
for both quantities in the literature.}

The faint end of the LF is of particular interest.  The standard cold dark 
matter (CDM) model predicts a steep faint end slope to the galaxy LF; the 
Press-Schechter approximation (1974) combined with a CDM-like power spectrum 
leads to an increasing number of dark halos as mass scales decrease 
\citep{white}.  A steep faint end slope has been seen in some environments, 
mainly those of high density \citep[e.g.,][]{trent1,phillips}.  In less extreme 
environments, however, flatter slopes are observed 
\citep[e.g.,][]{pritchet,trent3}.  Why are CDM-consistent slopes not seen in all 
environments?

Studies of the faint end of the LF are hampered by the extreme difficulty of 
obtaining complete samples of faint, low surface brightness galaxies.  It is 
always possible that the differences in the faint end slope of the LF are simply 
due to faint galaxies being missed in less dense regions.  Other explanations 
have also been suggested.  \citet{tully}, for example, use semi-analytic models 
to show that more low mass halos form earlier in regions that eventually become 
massive clusters.  They then point out that the reionization of the universe 
could inhibit gas collapse in low mass halos.  Combining these ideas, they 
conclude that dense regions start forming small, faint galaxies before 
reionization, and so can form many, while less dense regions begin small halo 
formation later, and so can only form few.

In this work, we examine the faint end of the LF in the core of the Coma 
Cluster.  Coma is a very rich cluster (Abell class 2);  Tully et al.'s model 
would therefore predict Coma to have formed many low mass halos before 
reionization.  We expect to find a steep faint end slope to the Coma LF.  We 
choose to use the method of statistical background subtraction \citep{zwicky} to 
remove background galaxy counts from our sample of cluster galaxy counts: we 
count the number of galaxies in the cluster direction and then subtract the 
count of galaxies in a ``blank sky'' control field.

Many Coma LFs have appeared in the literature (Table \ref{tab:cfothers}). This 
work differs from previous studies in that we use HST images of Coma and of our 
control field to determine the LF.  The increased depth and resolution of these 
images improves the determination of the faint end slope of the Coma LF.

\begin{deluxetable*}{c c c c c c} \tablewidth{0pt} \tabletypesize{\scriptsize} 
\tablecaption{Previous studies of Coma's luminosity function 
\label{tab:cfothers}} \tablehead{ \colhead{Study} & \colhead{Passband} & 
\colhead{Field size} & \colhead{Pixel scale} & \colhead{Lim. mag.} & 
\colhead{$\alpha$\tablenotemark{a}} \\ \colhead{} & \colhead{} & 
\colhead{(arcmin$^2$)} & \colhead{(arcsec pix$^{-1}$)} & \colhead{(Vega mags)} & 
\colhead{} }

\startdata Thompson \& Gregory 1993 & b & 14 292 & 18.56\tablenotemark{b} & 20 & 
-1.43 \\ Biviano et al. 1995 & b & 2 496 & 67.2\tablenotemark{b} & 20 & -1.2 
$\pm$ 0.2 \\ Bernstein et al. 1995 & R & 52.2 & 0.473 & 23.5\tablenotemark{c} & 
-1.42 $\pm$ 0.05 \\ Secker et al. 1997 & R & $\sim$ 700 & 0.53 & 22.5 & -1.41 
$\pm$ 0.05 \\ Lobo et al. 1997 & V & 1 500 & 0.3145 & 21 & -1.8 $\pm$ 0.05 \\ 
Trentham 1998b & R & 674 & 0.22 & 23.83 & -1.7 \\ Adami et al. 2000 & R & 52.2 & 
0.473 & 22.5 & -1 \\ Andreon \& Cuillandre 2002 & B & 720 & 0.206 & 22.5 & -1.25 
\\ \nodata & V & 1 044 & 0.206 & 23.75 & -1.4 \\ \nodata & R & 1 044 & 0.206 & 
23.25 & -1.4 \\ Beijersbergen et al. 2002 & U & 4 680 & 0.333 & 21.73 & 
$-1.32^{+0.018}_{-0.028}$ \\ \nodata & B & 18 720 & 0.333 & 21.73 & 
$-1.37^{+0.024}_{-0.016}$ \\ \nodata & r & 18 720 & 0.333 & 21.73 & 
$-1.16^{+0.012}_{-0.019}$ \\ Mobasher et al. 2003 & R & 3 600 & 0.21 & 19.5 & 
$-1.18^{+0.04}_{-0.02}$ \\ Iglesias-P{\'a}ramo et al. 2003 & r' & 3 600 & 0.333 
& 20.5 & $-1.47^{+0.08}_{-0.09}$ \\ \enddata \tablenotetext{a}{When no error is 
given for $\alpha$, no value was supplied in the text}

\tablenotetext{b}{These studies were done using photographic plates.  The value 
listed for ``Pixel scale'' is actually that for plate scale, measured in arcsec 
mm$^{-1}$}

\tablenotetext{c}{Bernstein et al. reported two limiting magnitudes for their 
work.  The first, shown here, is the faintest magnitude bin included in the fit 
to their luminosity function.  Beyond this, they believed their counts were 
accurate to R $\sim$ 25.5, but decided not to use the data for fear of globular 
cluster contamination.}

\end{deluxetable*}

\section{OBSERVATIONS AND DATA REDUCTION}

\subsection{Coma Data}

Our Coma data consist of observations obtained from the HST archive.  We 
selected 16 F606W exposures, totaling 20400 s, and 6 F814W exposures, totaling 
7800 s, of a single WFPC2 pointing taken with the center of the PC chip placed 
on the nucleus of NGC 4874. Due to light contamination from this galaxy, we 
discarded the PC chip from this project.  The combined area of the three WF 
chips is $\sim 16650$ arcsec$^2$ (after trimming), with a plate scale of 0.1$''$ 
pix$^{-1}$.  For further details of these observations, see \citet{jj}.

The raw data were processed with the standard HST pre-processing pipeline.  We 
then registered the images to within $\sim 0.2$ pix using $\sim 20$ stellar 
objects on each chip. We scaled the images based on their exposure time and 
applied a standard cosmic ray rejection algorithm.  Finally, we combined the 
images using the median pixel value, ending with coadded images with effective 
exposure times of 1300 s.

The coadded Coma images for each chip were dominated with light from a number of 
bright elliptical galaxies.  To remove this light, we used elliptical isophotes 
to construct photometric models of the six brightest galaxies on the three WF 
chips.  We then subtracted these models from the coadded images.  We trimmed the 
elliptical-subtracted coadded images to remove the regions affected by edge 
effects, and trimmed the ellipse model images to match.  We then ring 
median-filtered the elliptical-subtracted and trimmed images to create a 
smoothed map of the background light for each WF chip.  We subtracted these 
background light maps to create final background-subtracted images.  We also 
added the background light map to the trimmed ellipse model image for each chip 
to create final models of the background for each chip.

\subsection{Control Field Data}

For control field data, we obtained F606W and F814W images of the Hubble Deep 
Field North (HDF)  \citep{hdf} from the STScI website$^3$.  These images had 
been registered, drizzled onto an image with a sampling of 0.04$''$ pix$^{-1}$, 
background subtracted and normalized to an exposure time of 1 s.

\footnotetext[3]{\tt{http://www.stsci.edu/ftp/science/hdf/hdf.html}}

In the method of differential counts, it is imperative to recreate the detection 
characteristics of the data field as closely as possible in the control field 
(for a good discussion of this topic, see \citet{bern} and \citet{frenchie}).  
We imposed the Coma pixel scale on the HDF frames by rebinning the images by a 
factor of 2.5.  To match the Coma images' effective exposure time of 1300 s, we 
multiplied each HDF image by 1300.  We also trimmed the HDF images to match the 
area of the trimmed Coma images.

To match the noise of the Coma images, we generated a {\it spatially-dependent} 
noise frame for each chip.  At each pixel of the noise frame, the expected noise 
level of the corresponding Coma background model pixel was calculated based on 
the background level at that pixel, the read noise, and the number of frames 
that were coadded to create the image.  A value for the noise pixel was then 
drawn at random from a Gaussian distribution centered on zero with standard 
deviation equal to that expected noise value.  The generated value was 
multiplied by 0.85 {\it a priori} to better match the measured noise levels in 
the background-subtracted Coma images.  We added the resulting noise frames 
directly to the corresponding rebinned, scaled and trimmed HDF chips, as the 
existing Poisson noise in the HDF images is negligible.

Finally, we subtracted the background of the HDF images in the same manner as 
the Coma images to keep all processing steps as similar as possible.

\section{CREATING THE OBJECT CATALOGS}

\subsection{Globular Cluster Contamination}

One of the advantages of using HST data is its superior resolution over 
ground-based data.  This is important in this work because of the danger of 
globular cluster blends contaminating the LF.  Globular clusters are point 
sources at the distance of Coma, but two or more globular clusters (or other 
stellar objects) can appear to be a single extended object when close enough 
together, and thus be misclassified as a cluster galaxy.  This effect was 
discussed by \citet{frenchie}, who noted that a large number of faint, extended 
objects in their CFHT images of the Coma cluster were resolved into separate 
point sources in corresponding HST images.

\begin{figure}  \plotone{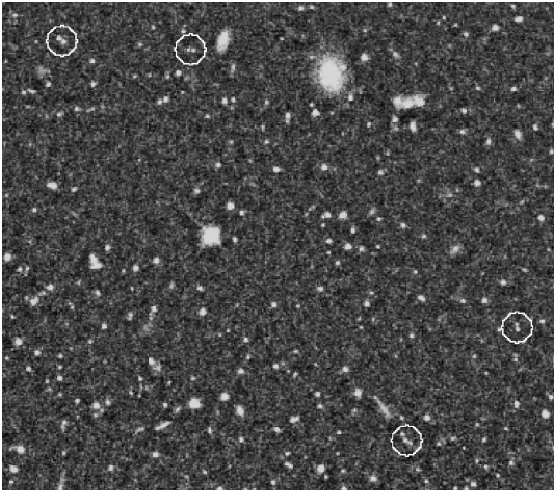} 

\caption{Blended objects in the Coma catalog.  The pairs of point sources 
circled above are easily resolvable by eye, but were blended into a single 
extended object by our detection algorithm}

\label{fig:blends} \end{figure}

Examination of our preliminary catalogs revealed that while globular clusters 
were certainly {\it visible} as separate point sources, these separate sources 
would still sometimes be blended into a single object by our detection 
algorithm.  We found this to be a function of the choice of convolution kernel; 
when the image was convolved with a moderate-sized Gaussian kernel (FWHM $\sim 
3$ pix) to enhance detection of extended sources, nearby point sources could be 
blended into a single extended object (Figure \ref{fig:blends}).  Convolving 
with a smaller kernel (FWHM $\sim 1.5$ pix) prevented this effect, but using 
such a small kernel caused some of the more extended faint galaxies to be 
missed.

To prevent globular cluster blends from contaminating our catalog while still 
including as many cluster galaxies as possible, we chose to employ a two-pass 
detection and photometry procedure.  We convolved the images once with a small 
kernel to detect every point source as a separate object.  We separated the 
resulting catalog into ``stars'' and galaxies on the basis of central 
concentration, and created a stellar mask for each chip.  We applied this mask 
in a second pass with a larger kernel, such that pixels associated with stellar 
objects in the first pass could not be detected on the second pass.  The results 
of this second pass then formed our catalog.

The HDF has many fewer point sources than the Coma images, and therefore is much 
less affected by the danger of point source blends.  However, in the interest of 
processing the data and control images as identically as possible, we applied 
the same two-pass procedure to the HDF images as well.

\subsection{Detection and Photometry}

We used the SExtractor package \citep{sex} for detection and photometry.  We 
performed the first detection/photometry pass on the F606W images, convolving 
with a FWHM = 1.5 pix Gaussian kernel.  We set the detection threshold to 1.48 
in units of background standard deviation in the unconvolved image; this 
corresponds to a threshold of $\sim 3.5$ $\sigma$ on the convolved image.  We 
gave the final Coma background model to SExtractor as a variance-type weight map 
for both the Coma image and the corresponding HDF image.  SExtractor uses these 
weight maps, which describe the noise intensity at each pixel in units of 
relative weight or relative variance, to slightly adjust the threshold at each 
pixel to compensate for noise varying across the frame.

To identify the stellar objects in the detection lists, we calculated a 
``simplified'' Petrosian radius $r_{petros}$ for each object.  This is defined 
as the radius at which the function \begin{equation} \eta' = \frac{F(< r)}{r} 
\end{equation}

\noindent reaches its maximum, where $F(< r)$ is the flux within an aperture 
with radius $r$.  Note that $\eta'$ is proportional to signal-to-noise when an 
image is sky-noise dominated.  This measure is loosely based on the 
\citet{petros} radius, which has been used to determine central concentration 
for various purposes in numerous studies (see, for example, Strauss et al. 
2002).

\begin{figure} \plotone{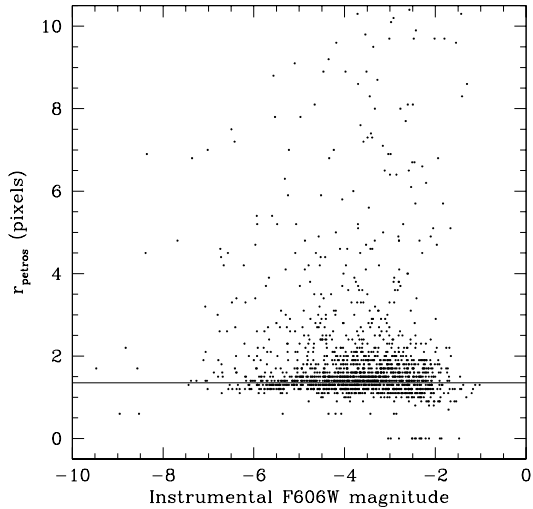}

\caption{Simplified Petrosian radius versus instrumental F606W magnitude 
for objects detected by convolving with a Gaussian kernel of FWHM = 1.5 
pix.  For the purposes of constructing the globular cluster mask, objects 
with $r_{petros} \leq 1.35$ are classified as stellar.  Some stars do fall 
above this line; however, a lower value is preferable at this step to 
block as many stars as possible while not losing any galaxies.}

\label{fig:stargal} \end{figure}

We plotted the $r_{petros}$ for each object versus its {\tt 
MAG\underline{~~}BEST} magnitude (Figure \ref{fig:stargal}).  We chose a cut of 
$r_{petros} = 1.35$ to separate stellar objects from galaxies.  Some stars do 
fall above this line; however, as the object of this step is to block as many 
stars as possible while not losing any galaxies, a lower value is preferable.  
We then used the segmentation image produced by SExtractor to identify the 
pixels associated with each stellar object.  We created a weight map where these 
pixels were assigned a weight of zero, while the remaining pixels received a 
weight equal to the inverse of the value of the corresponding Coma background 
model pixel.

We performed the second detection/photometry pass using this weight map.  The 
image was convolved with a FWHM = 3 pix Gaussian kernel, and we set the 
detection threshold to 0.86 (again equivalent to $\sim 3.5$ sigma on the 
convolved image).  We set the weight threshold such that no pixels with weights 
of zero would be included in the detected objects.  Photometry was performed on 
both the F606W and F814W images; object positions and image moments were 
determined on the (deeper) F606W frames, and then applied to both the F606W and 
F814W frames to calculate magnitudes.  Fixed-aperture magnitudes were also 
determined in both filters for later use in calculating colors.

\subsection{Galaxy Selection}

We trimmed the preliminary Coma catalogs to remove objects which fell in the 
cores of the subtracted Coma elliptical galaxies. The HDF had no large 
elliptical galaxies to subtract; however, its noise was generated from the Coma 
background models containing these galaxies, and so we removed HDF objects in 
the trim regions from the catalogs as well.  Finally, to remove any remaining 
stellar objects from the catalogs, we cut all objects with $r_{petros} \leq 
1.55$.

\subsection{Limiting Magnitude}

We performed add-galaxy experiments to determine the limiting magnitude of our 
catalogs.  For each of 12 0.5 magnitude bins, we generated 100 fake galaxies and 
added them to one chip of the Coma and HDF images.  After running SExtractor on 
the added-object images, we determined the number of added objects that were 
recovered.  We repeated this process 20 times for both Coma and the HDF and 
combined the results.

\begin{figure} \plotone{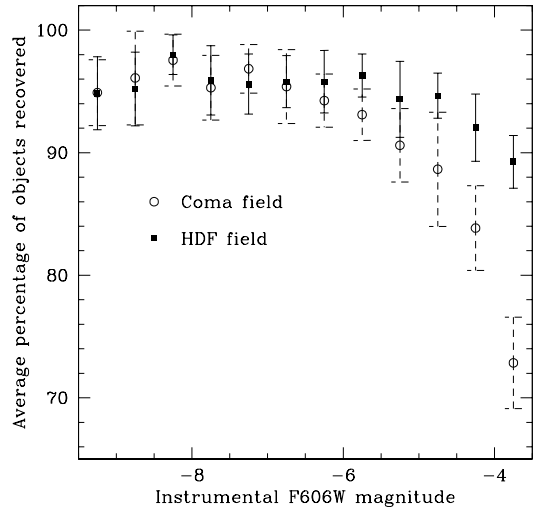} \caption{Average percentage of added objects 
recovered in each magnitude bin.  Open circles represent the results of objects 
added to the Coma field; filled squares are the results of objects added to the 
HDF field.  Each point represents the average of 20 runs.  Error bars are the 
standard deviation in the mean.} \label{fig:ndet.ndet} \end{figure}

Figure \ref{fig:ndet.ndet} shows the results for both Coma and the HDF.  As can 
be seen, the Coma catalog is over 80\% complete to an instrumental F606W 
magnitude of -4.25.  This can then be considered the limiting instrumental 
magnitude of the catalog.

\subsection{Magnitude Transformation}

Finally, we transformed our magnitudes to the Vega system Johnson-Cousins B, R 
and I bands.  To transform from instrumental magnitudes to the Vega system, we 
used the HDF photometric zeropoints on the STScI webpage$^4$;  to transform from 
F606W to B and R, and from F814W to I, we used the data of \citet{fuku}.  We 
determined the F606W - F814W color necessary for this transformation from the 
fixed-aperture magnitudes calculated by SExtractor.  To those objects for which 
SExtractor could not determine fixed-aperture magnitudes, we assigned a color of 
F606W - F814W = 0.8, chosen as the typical color from a histogram of object 
colors in the final Coma catalog.  Under this transformation, our limiting 
magnitude (80\% completeness) becomes R = 25.75 in the Vega system.

\footnotetext[4]{\tt{http://www.stsci.edu/ftp/science/hdf/logs/zeropoints.txt}}

\subsection{Color and Morphology}

The method of statistical background subtraction can be improved when combined 
with other forms of background subtraction.  A common technique is to pre-select 
cluster members based on morphology or color before performing the statistical 
background subtraction.  This reduces the number of background galaxies that 
must be removed statistically, therefore reducing the errors inherent in 
statistical subtraction.

Unfortunately, the objects found in our Coma fields are too small to derive 
meaningful morphology information.  The objects in the final Coma catalog have a 
typical effective radius of $\sim 2$ pix.  This is simply too few pixels to 
determine morphological type.

\begin{figure} \plotone{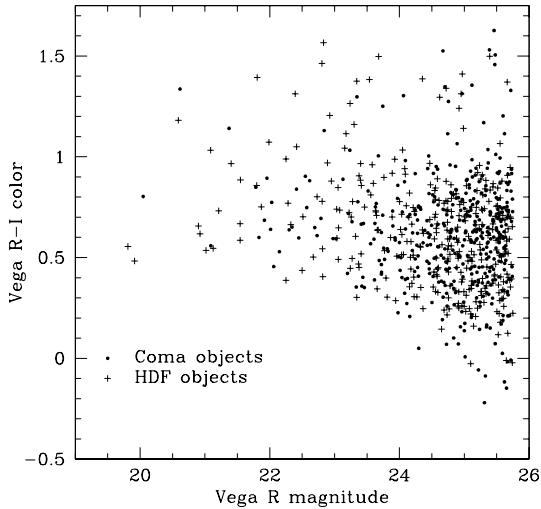} \caption{Color-magnitude diagrams of the 
final Coma and HDF catalogs.  Filled circles represent objects from the 
Coma catalog; crosses represent objects from the HDF catalog.} 
\label{fig:cmd} \end{figure}

Pre-selecting based on color also proved impossible with these data.  Figure 
\ref{fig:cmd} shows color-magnitude diagrams of objects in the final Coma and 
HDF catalogs.  As can be seen, the distributions are very similar.  There is no 
obvious way to select Coma cluster members based on their color.

\subsection{Cosmic Variance}

Cosmic variance - the fact that background counts vary from pointing to pointing 
due to clustering and large scale structure - must be considered when using 
statistical background subtraction.  To understand the effects of cosmic 
variance on the luminosity function, we evaluated the expression of 
\citet{peebles} for the variance of the count $N$ of objects in a randomly 
placed cell.  This expression is given by \begin{equation} \label{eqn:variance} 
\langle (N - \eta\Omega)^2 \rangle = \eta\Omega + \eta^2 \int d\Omega_1 
d\Omega_2 w(\theta_{12}) \end{equation}

\noindent where $\eta$ is the mean density of objects on the sky, $\Omega$ is 
the size of the cell, and $w(\theta_{12})$ is the two-point angular correlation 
function.

The two-point angular correlation function of galaxies is usually parameterized 
by $w(\theta_{12}) = A_w \theta^{-\delta}$, where $A_w$ is often a function of 
magnitude.  A search of the literature yielded no applicable studies of the 
two-point galaxy angular correlation function in the Vega R band.  We chose 
therefore to use the results of \citet{wilson}, who did her work in the Vega V 
band.  Wilson's data can be applied to this work if a constant Vega V-R color is 
assumed for the objects in the Coma and HDF catalogs.  As discussed earlier, the 
typical Vega F606W-F814W color of objects in the final Coma catalog is 0.8.  
Using the transformations of \citet{fuku}, this corresponds to a typical Vega 
V-R color of 0.5.  Therefore, Wilson's coefficients for the bin $m_1 < V \leq 
m_2$ were applied to the bin $m_1-0.5 < R \leq m_2-0.5$ in this work.  The 
values used can be found in Table \ref{tab:cv}

For $\eta$, we determined the mean density of background galaxies from the 
compilation of published field number counts available from the Durham Cosmology 
Group$^5$.  Because counts in the Vega V band were not available, the counts in 
the Vega R band were used.  Following the reasoning used above, the counts in 
the bin $m_1-0.5 < R \leq m_2-0.5$ were used as the counts in the bin $m_1 < V 
\leq m_2$.  From the given data, the average count in each 0.5 mag bin was 
determined.  This value was divided by 3600 to get the average number of 
galaxies per arcmin$^2$ in each bin.  As the cosmic variance was to be 
calculated in 1.0 mag bins, the results from pairs of bins were added together.  
The resulting values can be found in Table \ref{tab:cv}.

\footnotetext[5]{\tt{http://star-www.dur.ac.uk/~nm/pubhtml/counts/counts.html}}

Finally, for $\Omega$, we simply calculated the total field size of this work by 
adding together the trimmed areas of the three WF chips.  This value can also be 
found in Table \ref{tab:cv}

Using these values, we then evaluated Peebles' integral numerically for each of 
the four 1.0 magnitude bins.  The distance between points was calculated as the 
linear distance; this approximation is valid for fields as small as the one used 
here.  The resulting values of cosmic variance and standard deviation for each 
of the bins can be found in Table \ref{tab:cv}.

\begin{deluxetable*}{c c c c c c c c} \tablewidth{0pt} 
\tabletypesize{\scriptsize} \tablecaption{Cosmic Variance \label{tab:cv}} 
\tablehead{ \colhead{Vega R bin} & \colhead{Vega V bin} & \colhead{$\delta$} & 
\colhead{log$_{10} A_w(1')$} & \colhead{$\eta$} & \colhead{$\Omega$} & 
\colhead{Cosmic} & \colhead{Standard} \\ \colhead{} & \colhead{} & \colhead{} & 
\colhead{} & \colhead{(arcmin$^{-2}$ mag$^{-1}$)} & \colhead{(arcmin$^2$)} & 
\colhead{Variance} & \colhead{Deviation} } \startdata 20.5 - 21.5 & 21.0 - 22.0 
& 0.8 & -1.13 $\pm$ 0.06 & 1.15 & 4.62 & 7.72 & 2.78 \\ 21.5 - 22.5 & 22.0 - 
23.0 & 0.8 & -1.49 $\pm$ 0.05 & 2.69 & 4.62 & 21.3 & 4.62 \\ 22.5 - 23.5 & 23.0 
- 24.0 & 0.8 & -1.72 $\pm$ 0.05 & 6.37 & 4.62 & 48.4 & 6.96 \\ 23.5 - 24.5 & 
24.0 - 25.0 & 0.8 & -2.26 $\pm$ 0.09 & 14.6 & 4.62 & 96.2 & 9.81 \\ \enddata 
\end{deluxetable*}

To convert the cosmic variance to an error in the luminosity function, two steps 
were necessary.  First, as the luminosity function was to be expressed in terms 
of 0.5 mag bins, the error in each 1.0 mag bin due to cosmic variance had to be 
divided between two bins.  We chose to do this based on the relative population 
of the two bins:  If $\sigma_T$ is the standard deviation due to cosmic variance 
for a 1.0 mag bin, and $n_1$ and $n_2$ are the background counts in two 0.5 mag 
bins, then the standard deviation due to cosmic variance in the two 0.5 mag bins 
is given by \begin{eqnarray} \label{eqn:cverr} \sigma_1 &=& \sigma_T \frac {n_1} 
{n_1 + n_2} \\ \sigma_2 &=& \sigma_T \frac {n_2} {n_1 + n_2} \end{eqnarray}

Second, the error due to cosmic variance only applies to background counts.  
For the HDF, all the counts in each bin are due to background galaxies.  For 
Coma, however, only a fraction of the counts in each bin are from background 
galaxies.  The exact number for each bin is unknown; statistically, it is 
assumed to be the same as the HDF count for the corresponding bin.  Therefore, 
we set the error due to cosmic variance for each Coma bin equal to the error due 
to cosmic variance for the corresponding HDF bin.

We are not able to give an estimate of the cosmic variance all the way down to our 
limiting magnitude, because Wilson's study does not go this deep.  Therefore, we are 
forced to assume that the error due to cosmic variance remains small.  Further, by using 
Wilson's work, we are assuming that the conditions of her study - instrument
sensitivity, seeing, detection and photometry algorithm, and so on - are sufficiently 
similar to the conditions of this study to warrant the comparison.

Obviously, these are very strong assumptions.  Ideally, we would evaluate cosmic variance 
by obtaining additional data taken and reduced under the same conditions of the Coma data.  
However, this is beyond the scope of this paper, and so we continue under the assumptions 
stated above.

\section{RESULTS}

\subsection{Comparison to Other Dwarfs}

To compare the objects in our Coma catalog to known dwarf galaxies, we collected 
six published catalogs of dwarf galaxies (Table \ref{tab:cfsize}) and obtained 
the magnitude and half-light radius for each of the galaxies in physical units.  
When a distance to the group or galaxy was given with the catalog, we adopted 
that value.  For the Virgo cluster, we followed \citet{jj} and used a distance 
modulus of (m-M) = 30.99.

\begin{deluxetable}{c c c} \tablewidth{0pt} \tablecaption{Published catalogs of 
dwarf galaxies \label{tab:cfsize}} \tablehead{ \colhead{Catalog} & 
\colhead{Number of galaxies} & \colhead{Reference} } \startdata Local Group & 21 
& Bender et al. 1992 \\ M101 & 18 & Bremnes et al. 1999 \\ M81 & 19 & Bremnes et 
al. 1998 \\ Northern Field & 17 & Barazza et al. 2001 \\ Southern Field & 24 & 
Parodi et al. 2002 \\ Virgo Cluster & 365 & Binggeli \& Cameron 1993 \\ \enddata 
\end{deluxetable}

We converted the magnitudes and half-light radii of objects in our final Coma 
catalog into physical units, adopting a distance modulus of (m-M) = 35.05 
\citep{jj}.  For comparison, we also determined magnitudes and half-light radii 
for Coma's globular clusters (i.e., the objects found to be stellar while 
constructing the globular cluster mask).

\begin{figure*} 
\plotone{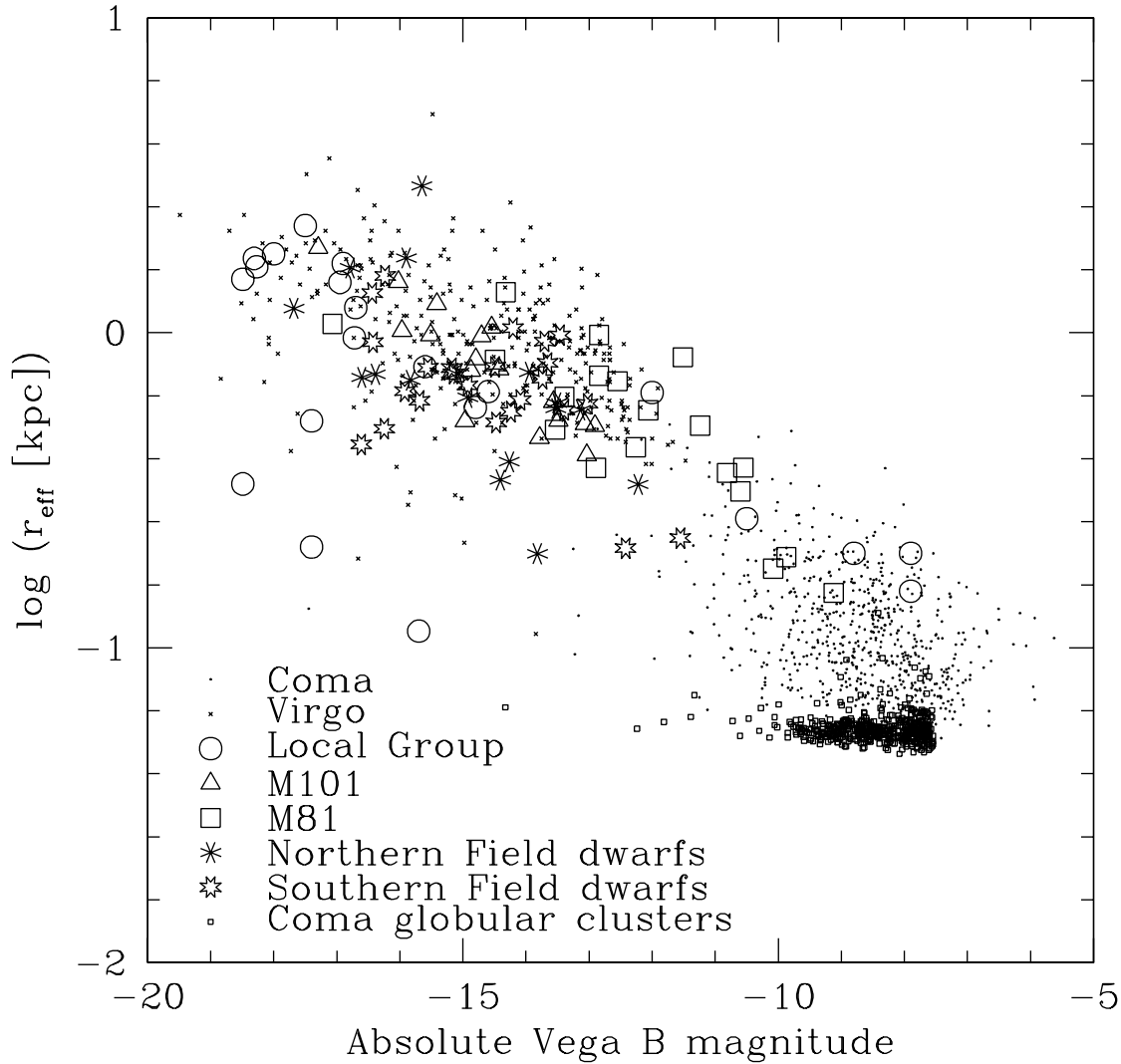} 
\caption{Logarithm of effective radius in kpc versus absolute Vega B magnitude 
for various dwarfs.  Data for dwarfs in Coma are from this work; references for 
the other catalogs used can be found in Table \ref{tab:cfsize}.  Also shown for 
comparison are the Coma globular clusters identified in this work.} 
\label{fig:cfsize} \end{figure*}

Figure \ref{fig:cfsize} shows half-light radius as a function of magnitude for 
all the objects.  Although the Coma galaxies are much fainter than most known 
dwarf galaxies, they fall in the same general locus.  The distribution is 
understandably more scattered: background galaxies are also present in the Coma 
catalog, and these show a range of half-light radii at each magnitude.  The Coma 
globular clusters, however, clearly occupy a very different region in the plot.  
From this plot, we conclude that the objects in the Coma catalog are {\it not} 
globular clusters, and do have the same general morphology as known dwarf 
galaxies.

\subsection{The Luminosity Function}

Table \ref{tab:lf} shows the data that were used to construct the LF.  The 
number of galaxies in the HDF image is scaled to the unblocked area of the Coma 
image; the Coma image has many more point sources, and therefore more pixels are 
blocked by the stellar mask.

Poisson counting errors and errors due to cosmic variance are given separately.  
The error due to cosmic variance in each 0.5 mag bin could be added in 
quadrature to the Poisson error.  However, the calculation here was performed 
only to serve as an estimate of the effects of cosmic variance on the slope of 
the luminosity function.  A full treatment would have to account for the 
correlations in error between magnitude bins introduced by cosmic variance.  
Due to the simplistic approach and many assumptions used here, we choose to 
present the two errors separately.

\begin{deluxetable}{c c c c c c} \tablewidth{0pt} \tablecaption{The luminosity 
function \label{tab:lf}} \tablehead{ \colhead{$m_R$} & \colhead{$N$} & 
\colhead{$\Delta N_p$} & \colhead{$\Delta N_{CV}$} & \colhead{log $\sigma$} & 
\colhead{$\Delta$log $\sigma$} } \startdata \cutinhead{Coma} 20.25 & 1.00 & 1.00 
& \nodata & 2.89 & 2.89 \\ 20.75 & 1.00 & 1.00 & 1.04 & 2.89 & 2.89 \\ 21.25 & 
2.00 & 1.41 & 1.74 & 3.19 & 3.04 \\ 21.75 & 4.00 & 2.00 & 2.31 & 3.49 & 3.19 \\ 
22.25 & 8.00 & 2.83 & 2.31 & 3.79 & 3.34 \\ 22.75 & 8.00 & 2.83 & 2.07 & 3.79 & 
3.34 \\ 23.25 & 14.00 & 3.74 & 4.89 & 4.04 & 3.46 \\ 23.75 & 27.00 & 5.20 & 3.68 
& 4.32 & 3.61 \\ 24.25 & 50.00 & 7.07 & 6.13 & 4.59 & 3.74 \\ 24.75 & 80.00 & 
8.94 & \nodata & 4.79 & 3.84 \\ 25.25 & 134.00 & 11.58 & \nodata & 5.02 & 3.95 
\\ 25.75 & 203.00 & 14.25 & \nodata & 5.20 & 4.04 \\ \cutinhead{Scaled HDF} 
20.25 & 1.97 & 1.39 & \nodata & 3.18 & 3.03 \\ 20.75 & 2.96 & 1.71 & 1.04 & 3.36 
& 3.12 \\ 21.25 & 4.92 & 2.20 & 1.74 & 3.58 & 3.23 \\ 21.75 & 6.90 & 2.61 & 2.31 
& 3.73 & 3.31 \\ 22.25 & 6.91 & 2.61 & 2.31 & 3.73 & 3.31 \\ 22.75 & 10.81 & 
3.26 & 2.07 & 3.93 & 3.40 \\ 23.25 & 25.60 & 5.02 & 4.89 & 4.30 & 3.59 \\ 23.75 
& 20.68 & 4.51 & 3.68 & 4.21 & 3.55 \\ 24.25 & 34.45 & 5.82 & 6.13 & 4.43 & 3.66 
\\ 24.75 & 78.75 & 8.81 & \nodata & 4.79 & 3.84 \\ 25.25 & 70.91 & 8.36 & 
\nodata & 4.74 & 3.81 \\ 25.75 & 107.32 & 10.28 & \nodata & 4.92 & 3.90 \\ 
\cutinhead{Coma - Scaled HDF} 20.25 & -0.97 & 1.71 & \nodata & \nodata & \nodata 
\\ 20.75 & -1.96 & 1.98 & 1.48 & \nodata & \nodata \\ 21.25 & -2.92 & 2.62 & 
2.46 & \nodata & \nodata \\ 21.75 & -2.90 & 3.29 & 3.27 & \nodata & \nodata \\ 
22.25 & 1.09 & 3.85 & 3.27 & 2.93 & 3.48 \\ 22.75 & -2.81 & 4.32 & 2.92 & 
\nodata & \nodata \\ 23.25 & -11.60 & 6.26 & 6.92 & \nodata & \nodata \\ 23.75 & 
6.32 & 6.88 & 5.20 & 3.69 & 3.73 \\ 24.25 & 15.55 & 9.16 & 8.67 & 4.08 & 3.85 \\ 
24.75 & 1.25 & 12.55 & \nodata & 2.99 & 3.99 \\ 25.25 & 63.09 & 14.28 & \nodata 
& 4.69 & 4.05 \\ 25.75 & 95.68 & 17.57 & \nodata & 4.87 & 4.14 \\ \enddata

\tablecomments{The columns are: $m_R$ -- midpoint of bin in apparent Vega R 
magnitudes; $N$ -- number of galaxies; $\Delta N_p$ -- Poisson error in the 
count; $\Delta N_{CV}$ -- error due to cosmic variance in the count; log 
$\sigma$ -- logarithmic surface density of objects (number of galaxies per 
deg$^2$); $\Delta$log $\sigma$ -- Poisson error in log $\sigma$.}

\end{deluxetable}

\begin{figure} \plotone{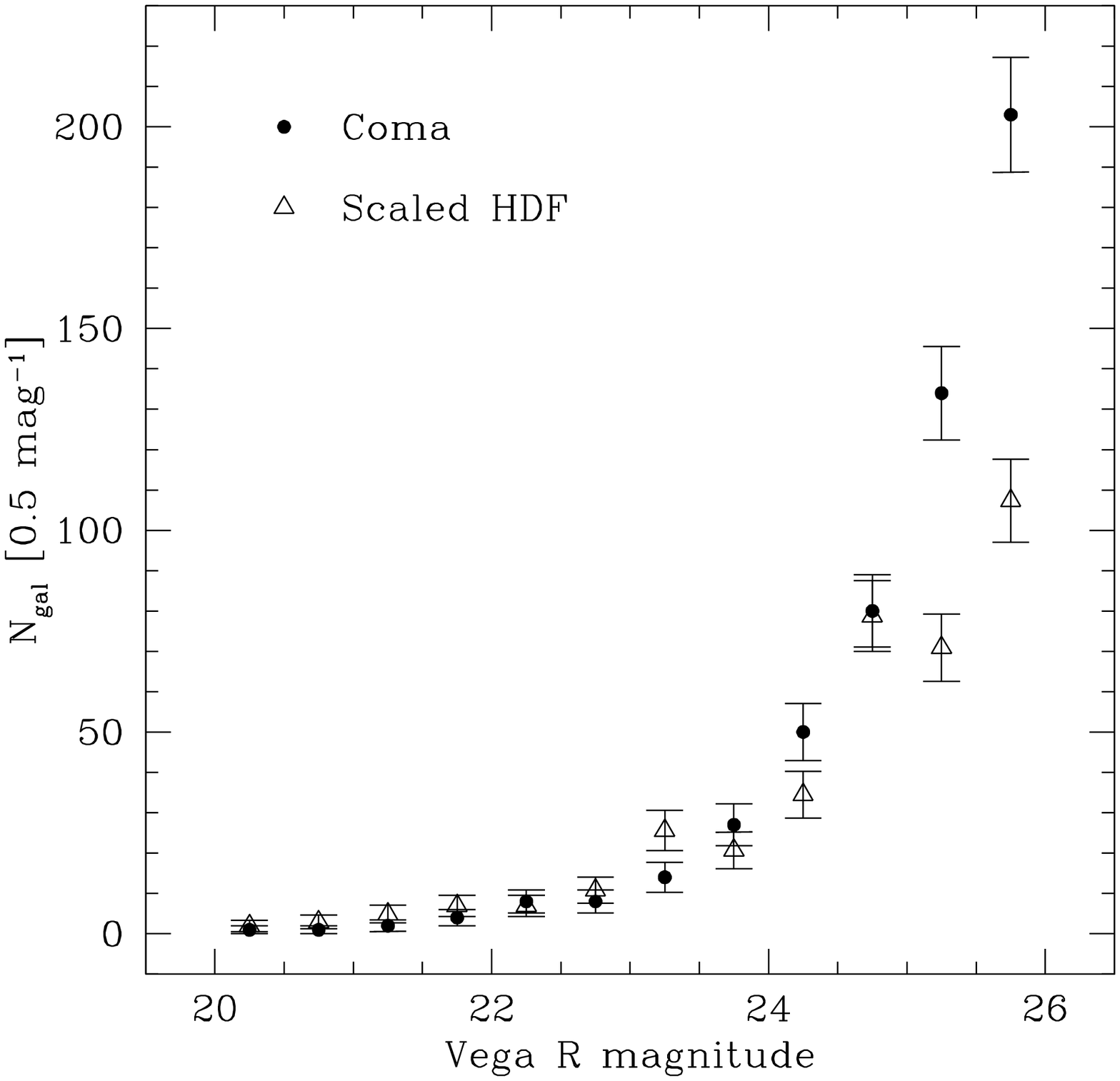} \caption{Number counts from the Coma cluster 
field and the HDF field (after scaling).  Error bars show the Poisson error in 
the counts.} \label{fig:counts} \end{figure}

Figure \ref{fig:counts} shows the number counts of the Coma field and the scaled 
HDF field.  As can be seen, cluster counts do not start to dominate over the 
background until the last two bins.  This is due to the small field size 
surveyed here: cluster galaxies with $20 \lesssim R \lesssim 24$ are so rare 
that $\lesssim$ 1 is expected in a field this size.  The galaxies in this 
magnitude range that are detected in the Coma field are due to background 
contamination, and so their numbers match those found in the background-only 
field.

\begin{figure} \plotone{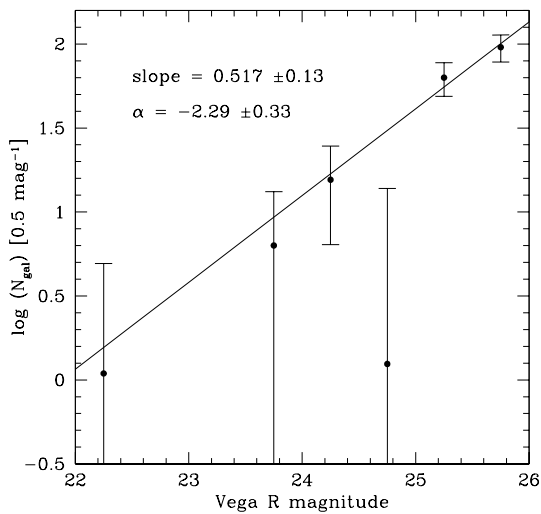} \caption{Vega R band luminosity function of 
the 
Coma cluster, expressed as logarithmic counts per 0.5 magnitude bin.  Error bars 
show the Poisson error in the counts.  The line is a weighted least squares fit 
to the points.} \label{fig:loglf} \end{figure}

Figure \ref{fig:loglf} shows the result of subtracting the HDF counts from the 
Coma counts -- the luminosity function.  The shape of the luminosity function is 
generally parameterized by the Schechter function, which can be written in terms 
of magnitude as \begin{equation} \phi (M) dM \sim 10^{-0.4(\alpha + 1) M} 
e^{-10^{0.4(M^* - M)}} \end{equation}

\noindent At faint magnitudes, the first term dominates and so the slope of the 
logarithmic luminosity function is directly related to the parameter $\alpha$.  
Fitting a straight line to our logarithmic LF using a weighted least-squares 
fit, we obtain $\alpha = -2.29 \pm 0.33$.

\section{DISCUSSION}

\subsection{Comparison to Other Work}

We have found an extremely steep slope for the faint end of the luminosity 
function in the Coma cluster.  This is especially striking when compared to 
studies of the faint-end LF in clusters such as Virgo and Fornax.  These 
clusters are close enough that measurements can easily be made at the same 
absolute magnitudes reached here.  For example, in the Virgo cluster, 
\citet{trenthodge} found magnitude-dependent slopes ranging from $-1.49 \pm 0.11 
\leq \alpha \leq -1.02 \pm 0.09$ for $-15 \leq M_B \leq -12$ (roughly equivalent 
to our absolute magnitude range of $-13 \lesssim M_R \lesssim -9$).  
\citet{sab} found a slightly steeper slope of $\alpha = -1.6$ for Virgo in the 
range $-14 \leq M_B \leq -10$, though this is still significantly flatter than 
our slope.  For Fornax, \citet{hilker} found an extremely flat slope of $\alpha 
= 1.1 \pm 0.1$ in the range $-16 \lesssim M_V \lesssim -9$.

Another interesting comparison is to look at other studies of the LF in the Coma 
cluster.  As noted in the Introduction, many other such studies have appeared in 
the literature.  Table \ref{tab:cfothers} indicates that little consensus has 
been reached on the ``true'' slope of the LF.  To better compare these studies, 
we constructed a composite LF.

We obtained background-subtracted number counts for the Coma cluster from the 
studies listed in Table \ref{tab:cfothers} that were conducted in the R band.  
We converted the magnitudes to absolute magnitudes using a distance modulus of 
$(m-M) = 34.83$ \citep{trent2}.  We then scaled the counts and errors to an area 
of 1 deg$^2$, using the field size reported by the authors.

No other normalization was performed.  This introduces some scatter to the 
composite LF; to be perfectly correct, larger surveys -- covering more of the 
low density cluster outskirts and hence having a lower average surface density 
of galaxies -- should be scaled differently than surveys covering only the high 
density cluster core.  However, taking the scale radius (i.e., the radius at 
which the slope of the profile is the average of the inner and outer slope) of 
the Coma cluster to be $r_{s} = 320$ kpc for H$_o$ = 70 km s$^{-1}$ Mpc$^{-1}$ 
\citep{lokas}, the largest survey included here \citep{mobasher} covers an area 
roughly 9 times that of the Coma cluster core.  Assuming a standard surface 
density profile ($\sigma(r) = (1+(r/r_{s})^n)^{-1}$, for $n=1$ or $2$), moving 
to a radius $\sim 3r_{s}$ would roughly halve the average surface density 
determined from observing only the cluster core.  This corresponds to a scatter 
of $\pm 0.3$ in our logarithmic composite LF, an acceptable level of error for 
our purposes.

\begin{figure*} \plotone{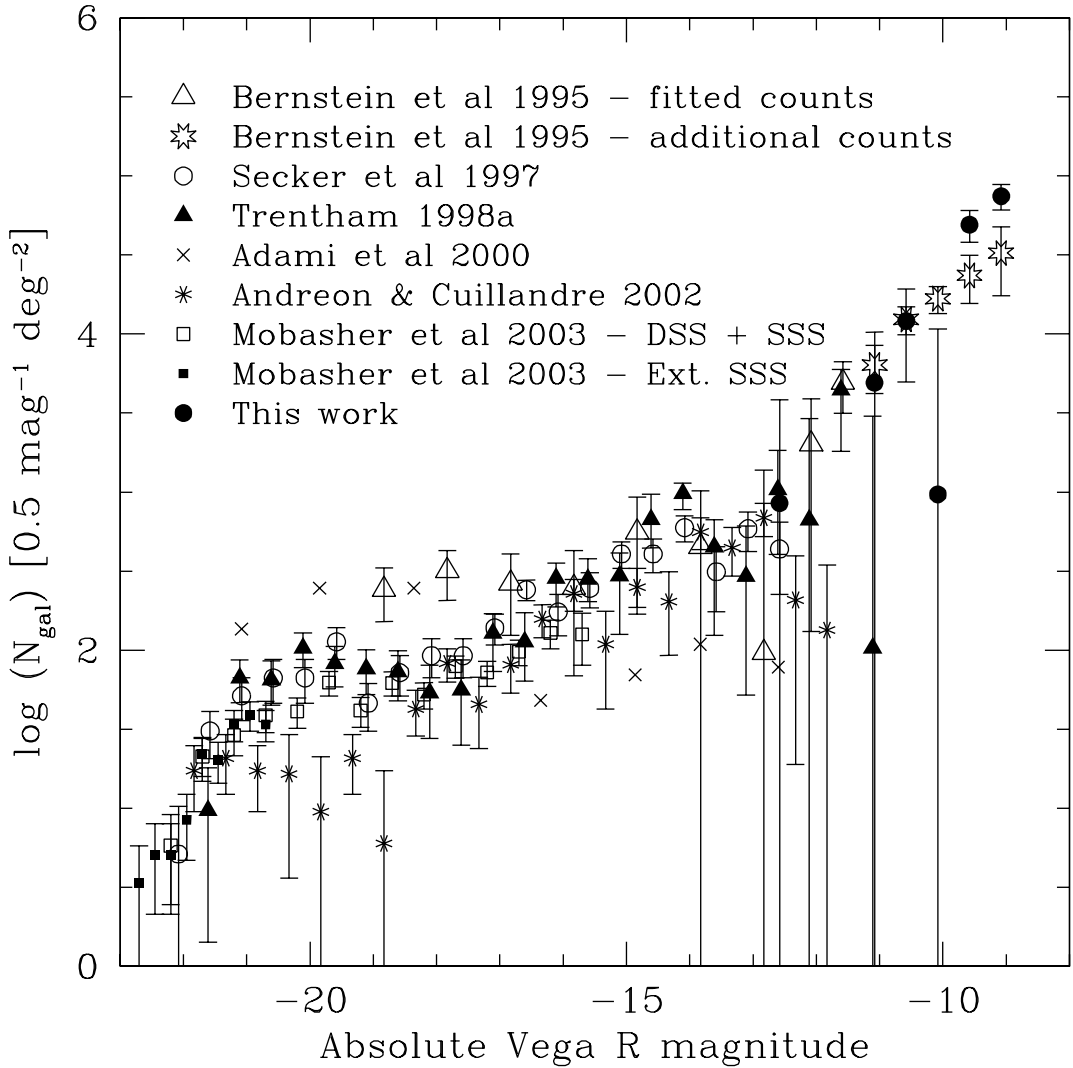} 
\caption{The 
composite Vega R band luminosity function of the Coma cluster, expressed as 
logarithmic counts per 0.5 magnitude bin and per deg$^2$.  Bernstein et al.'s 
``additional counts '' are those not included in their final luminosity function 
for fear of globular cluster contamination.  All counts are scaled to an area of 
1 deg$^2$; no other normalization was performed.} \label{fig:cflf} \end{figure*}

Figure \ref{fig:cflf} shows the composite R band luminosity function for the 
Coma cluster.  The bright end shows good agreement among the various surveys.  
At the faint end, the counts of this work are a clean continuation of the 
brighter counts from the literature and clearly show a steep faint end slope for 
the Coma cluster LF.

Especially interesting are the counts of \citet{mobasher} and \citet{adami}.  
Of the $R$ band studies discussed here, they were the only groups to use 
spectroscopic redshifts to determine cluster membership.  As can be seen in 
Table \ref{tab:cfothers}, their values for $\alpha$ are among the lowest 
published.  This trend of lower $\alpha$ values for spectroscopic studies has 
sometimes been interpreted as an indication that statistical background 
subtraction causes artificially steep slopes.

However, looking at the composite LF, the spectroscopic counts of Mobasher et 
al. agree very well with those obtained by using photometric methods to 
determine cluster membership.  Their counts stop before moving into the truly 
faint regimes of the LF; spectroscopic studies, by their nature, are limited to 
bright galaxies and so rarely probe the faint end of the LF.  The flatter slopes 
reported by spectroscopic studies may simply be a result of probing only the 
brighter, less steeply rising part of the LF.

The counts of Adami et al. do extend into fainter regions of the luminosity 
function.  Their study worked with a small sample of only 88 redshifts.  They 
did not calculate error bars for their luminosity function, but assumed them to 
be the same as those shown for the (much higher) counts of \citet{bern}.  With 
these error bars, their counts are consistent at a $1 \sigma$ level with those 
obtained using photometric methods.  The general trend of the points may hint at 
a possible systematic error arising from using photometric methods to construct 
a luminosity function, but it is dangerous to draw conclusions based on such 
small numbers and rough error estimates.  Unless a larger, better characterized 
sample of redshifts are obtained which show the same results, we see no 
compelling evidence that statistical background subtraction leads to an 
artificially steep LF slope.

Perhaps the most important thing that Figure \ref{fig:cflf} shows is that, {\it 
although the various studies did not agree on a value for $\alpha$, the counts 
themselves agree very well.} Caution must always be used when comparing results 
via parameterized values rather than the underlying data.

\subsection{Implications of a Steep Faint-End Slope}

We have found a steep faint end slope to the Coma LF.  Such steep slopes are not 
seen in all environments.  What could cause this difference between faint end 
slopes in different environments?

Qualitatively, our results agree with the idea of \citet{tully}: Coma, being a 
rich cluster, formed many dwarf galaxies before reionization ``squelched'' dwarf 
galaxy formation.  A more quantitative analysis of this idea was performed by 
\citet{benson}.  They used detailed semi-analytic models of galaxy formation, 
including the effects of supernova feedback, photoionization suppression, and 
dynamical friction and mass loss due to tidal forces, to determine the 
properties of galaxies in a range of environments.  They found that when 
photoionization suppression was switched on, faint end slopes in poor 
environments did become flatter than those in rich environments.  However, 
comparing their results to the observational data of \citet{trenthodge}, they 
found that photoionization alone cannot flatten the slope enough to match the 
luminosity functions observed in the Local Group and Ursa Major.

This conclusion is supported by \citet{grebel}, who examined the ages of stellar 
populations in Local Group dwarfs.  They found no evidence for a halt or 
reduction in star formation at $20 < z < 6.4$, the typical range of epochs for 
reionization.  This indicates that reionization cannot be the dominant influence 
in the evolution of Local Group dwarfs.  Other feedback mechanisms must also be 
at work.

Obviously, further theoretical efforts are needed to refine the exact processes 
that shape the luminosity function.  Based on the results of this and similar 
work, these studies must now also explain how these processes can lead to a 
luminosity function that is different in environments of different densities.

\section{CONCLUSION}

We have found a steep faint end slope for the galaxy luminosity function 
in the core of the Coma cluster.  Using the method of statistical 
background subtraction and archival HST images has allowed us to achieve a 
limiting magnitude of R = 25.75, making this the faintest and highest 
resolution determination of the Coma luminosity function to date.  The 
fine pixel scale of the WFPC2 camera and our globular cluster mask enabled 
us to effectively eliminate globular cluster contamination from the final 
catalog.  We find the slope of the luminosity function to be fit by 
$\alpha = -2.29 \pm 0.33$.  While this value is not in agreement with 
other published values of $\alpha$ for the Coma cluster, a composite 
luminosity function shows our counts to be in good agreement with previous 
counts.  This result could be affected if the errors due to cosmic 
variance are larger than our estimate.  Assuming they are not, the steep 
slope found in this work is consistent with theories that predict 
photoionization and other feedback effects will affect environments of low 
density more severely than environments of high density.

\section{ACKNOWLEDGMENTS}

This work was supported by the Natural Sciences and Engineering Research Council of Canada (NSERC)
through Discovery Grants to CJP and WEH.  This work was also 
supported in part by a 
Post-Graduate Scholarship to MLM from the NSERC.  MLM was a Guest User, Canadian Astronomy Data Centre, 
which is operated by the 
Herzberg Institute of Astrophysics, National Research Council of Canada.

\end{document}